\documentclass[12pt]{iopart}
\usepackage{iopams}
\usepackage{setstack}
\usepackage{psfig}

\begin{document}

\title[Reflectionless Potentials and $\cal PT$ Symmetry]{Reflectionless
Potentials and $\cal PT$ Symmetry}

\author[Ahmed, Bender and Berry]{Zafar Ahmed${}^1$, Carl~M~Bender${}^2$, and
M~V~Berry${}^3$}

\address{${}^1$Nuclear Physics Division, Bhabha Atomic Research Centre,
Trombay, \\ \quad Mumbai 400 085, India}

\address{${}^2$Department of Physics, Washington University, St. Louis MO 63130,
USA}

\address{${}^3$H.~H.~Wills Physics Laboratory, Tyndall Avenue, Bristol BS8 1TL,
UK}

\begin{abstract}
Large families of Hamiltonians that are non-Hermitian in the
conventional sense have been found to have all eigenvalues real, a
fact attributed to an unbroken $\cal PT$ symmetry. The
corresponding quantum theories possess an unconventional scalar
product. The eigenvalues are determined by differential equations
with boundary conditions imposed in wedges in the complex plane.
For a special class of such systems, it is possible to impose the
$\cal PT$-symmetric boundary conditions on the real axis, which
lies on the edges of the wedges. The $\cal P T$-symmetric spectrum
can then be obtained by imposing the more transparent requirement
that the potential be reflectionless.
\end{abstract}

\submitto{\JPA} 

The purpose of this letter is to point out a connection between
real reflectionless potentials and quantum mechanics defined by
complex $\cal P T$-symmetric non-Hermitian Hamiltonian operators.

Research begun in 1998 \cite{BB,BBM,DDT1,DDT2} has established
that the eigenvalues of the complex $\epsilon$-deformed anharmonic
oscillator
\begin{equation}
H=p^2+x^{2K}(ix)^\epsilon\quad(\epsilon>0,~K=1,\,2,\,3,\,\cdots).
\label{e1}
\end{equation}
are discrete and positive and real even though $H$ is
non-Hermitian $H^\dag\neq H$ in the conventional sense, where
$H^\dag$ denotes complex conjugate transpose. Even though
conventional Hermiticity has been replaced by the weaker condition
of $\cal PT$ symmetry, the Hamiltonian $H$ defines a consistent
quantum theory involving an unconventional scalar product
\cite{BBJ}. The Hamiltonian operator $H$ is defined by the
Schr\"odinger equation
\begin{eqnarray}
-\psi_n''(x)+[x^{2K}(ix)^\epsilon-E_n]\psi_n(x)=0. \label{e2}
\end{eqnarray}
The eigenfunction $\psi_n(x)$ satisfies the $\cal PT$-symmetric
boundary conditions that $\psi_n(x)$ vanishes as $|x|\to\infty$ in
two wedges symmetrically placed about the imaginary axis in the
lower-half $x$ plane. As explained in \cite{BT}, the wedges are
determined by analytic continuation based on the leading-order
exponentials in the asymptotic solutions of (\ref{e2}), namely
\begin{eqnarray}
\psi(x)\sim\exp\left(\pm {i^{\epsilon/2}x^{K+1+\epsilon/2}\over
K+1+\epsilon/2} \right). \label{e3}
\end{eqnarray}
Within each wedge, one of the two solutions decays and one grows.
Thus, the wedges are centred on the asymptotic directions
\begin{eqnarray}
\theta_{\rm right}=-{\epsilon\pi\over4K+2\epsilon+4},\qquad
\theta_{\rm left}=-\pi+{\epsilon\pi\over4K+2\epsilon+4}.
\label{e4}
\end{eqnarray}
The exponents in these asymptotic exponentials are purely real on
the Stokes lines \cite{D} at the centres of the wedges. It is easy
to check that
\begin{eqnarray}
&&{\rm in~the~right~wedge}\!:\quad\psi_-~{\rm
decays~and}~\psi_+~{\rm grows};
\nonumber\\
&&{\rm in~the~left~wedge}\!:\quad\left\{\begin{array}{l}
\psi_-~{\rm decays~and}~\psi_+~{\rm grows~if}~K~{\rm is~odd},\\
\psi_+~{\rm decays~and}~\psi_-~{\rm grows~if}~K~{\rm is~even}.
\end{array}\right.
\label{e5}
\end{eqnarray}
The opening angle of each wedge is
\begin{eqnarray}
\theta_{\rm opening~angle}={2\pi\over2K+\epsilon+2}. \label{e6}
\end{eqnarray}
The wedge boundaries are anti-Stokes lines \cite{D}, where the
solutions are purely oscillatory \cite{BB}.

We are concerned here with the infinite subclass $\epsilon=2$ for
which the potential in (\ref{e1}), that is
\begin{equation}
V(x)=-x^{2k}\quad(K=1,\,2,\,3,\,\cdots), \label{e7}
\end{equation}
is real and appears to have the wrong sign to possess bound
states. However, the $\cal PT$-symmetric solution to the
Schr\"odinger equation (\ref{e2}) with the complex boundary
conditions described above does have bound states.

Notice that when $\epsilon=2$, the upper edges of the right and
left wedges defined above (where the solutions are purely
oscillatory) lie exactly along the positive and negative real-$x$
axis. On the real axis the potential (\ref{e7}), when interpreted
in conventional terms, describes one-dimensional scattering
solutions of the Schr\"odinger equation, that is, travelling waves
rather than decaying exponentials at infinity. Our central point
here is that as far as the energy spectrum is concerned, we may
replace the non-Hermitian eigenvalue problem in the complex wedges
by a conventional Hermitian problem defined by the requirement
that the potential be reflectionless. (To be precise, we are {\it
not} claiming that these two quantum theories, the non-Hermitian
$\cal P T$-symmetric theory defined in the complex wedges and the
Hermitian theory defined on the real axis, are the same. However,
these two distinct theories do have the same energy spectrum and
eigenfunctions.)

To justify the assertion that we may replace the $\cal
PT$-symmetric theory by the Hermitian one, consider the two
dominant exponentials (\ref{e3}) when $\epsilon=2$:
\begin{equation}
\psi_\pm(x)\sim\exp\left(\pm i {x^{K+2}\over K+2}\right).
\label{e8}
\end{equation}
For real $x$ these behaviours represent waves travelling in
directions given by the sign of the current ${\rm
Im}\,\left(\psi_\pm^*\psi_\pm'\right)$, which is proportional to
$\pm x^{K+1}$. Thus,
\begin{eqnarray}
&&{\rm when}~x>0\!:\nonumber\\
&&\quad\!\!\psi_+(x)~{\rm travels~to~the~right~and}~\psi_-(x)~{\rm
travels~to~the~left};\nonumber\\
&& {\rm when}~x<0\!:\nonumber\\
&&\quad\!\!\left\{\!\!
\begin{array}{l}
\psi_+(x)~{\rm travels~to~the~right~and}~\psi_-(x)~{\rm
travels~to~the~left}~(K~{\rm odd}),\\
\psi_-(x)~{\rm travels~to~the~right~and}~\psi_+(x)~{\rm
travels~to~the~left}~(K~{\rm even}).
\end{array}\right.
\label{e9}
\end{eqnarray}
These conditions match those for decay and growth in (\ref{e5}),
so decay in the non-Hermitian problem corresponds to a purely
left-travelling wave that is reflectionless in the corresponding
conventional Hermitian problem. Under $\cal PT$ reflection, that
is, replacing $x\to -x$ and $i\to-i$ in (\ref{e8}), both wave
directions reverse if $K$ is even and do not reverse if $K$ is
odd. Reflectionlessness persists in both cases.

The corresponding energies can be approximated by the WKB method,
starting with identification of the two turning points, defined by
$V(x)=E$, in the wedges described above. The turning points are
\begin{equation}
x_{\rm right}=E^{1/(2K+2)}e^{-i\pi/(2K+2)}\quad{\rm and}\quad
x_{\rm left}=E^{1/(2K+2)}e^{-i\pi+i\pi/(2K+2)}. \label{e10}
\end{equation}
Quantization for large $n$ according to
\begin{equation}
\int_{x_{\rm left}}^{x_{\rm
right}}dt\sqrt{E-V(t)}=\left(n+{1\over2}\right)\pi \label{e11}
\end{equation}
leads straightforwardly \cite{BM} to
\begin{equation}
E_n\sim\left(\frac{(n+1/2)\sqrt{\pi}(K+2)\Gamma[(K+2)/(2K+2)]}{\Gamma[1/(2K+2)
]\cos[\pi/(2K+2)]}\right)^{(2K+2)/(K+2)}. \label{e12}
\end{equation}

Reflectionlessness can be regarded as the consequence of
destructive interference between exponentially small waves
reflected separately from the turning points $x_{\rm right}$ and
$x_{\rm left}$ in (\ref{e10}). A similar interference phenomenon
occurs in the analogous problem of exponentially weak adiabatic
transition probabilities, which can also vanish if there are
interfering contributions from complex turning points \cite{L}. In
the present example, as in some others considered recently
\cite{A}, the vanishing reflection requires more than one pair of
complex turning points and the corresponding potentials have
vanishing curvature at the top of the potential at $x=0$.

Reflectionlessness can also arise for other reasons. With
potential barriers of finite range, for example, the interfering
contributions come from nonanalytic extremities of the potential
\cite{B}. Alternatively, the cancellation can arise from a cluster
of turning points which (unlike those considered here) cannot be
separated \cite{BH}, in which case reflectionlessness can persist
for a continuum of parameter values rather than for isolated
energies, such as those in (\ref{e12}).

In conclusion, we have shown that for an infinite class of
non-Hermitian $\cal P T$-symmetric Hamiltonians having unbroken
$\cal PT$ symmetry, and thus real positive discrete spectra, there
is a corresponding set of Hermitian Hamiltonians whose spectra and
eigenfunctions become identical when the condition of
reflectionlessness is imposed. Although these pairs of
Hamiltonians do not describe the same physics because the inner
product needed to calculate matrix elements is different, this
intriguing association between non-Hermitian and Hermitian
Hamiltonians may help to explain the surprising observation
\cite{BB} that the spectra of some non-Hermitian Hamiltonians can
be real. We do not yet know if the association described in this
paper extends to more general classes of $\cal PT$-symmetric
Hamiltonians.

The connection between reflectionless potentials and $\cal PT$
symmetry may find application in quantum cosmology. Recently, much
attention has been given to Anti-de Sitter cosmologies \cite{ADS}
and de Sitter cosmologies \cite{DS1,DS2}. In the AdS description
the universe propagates reflectionlessly in the presence of a
wrong-sign potential ($-x^6$, for example). In the dS case the
usual Hermitian quantum mechanics must be abandoned and be
replaced by a non-Hermitian one in which there are
`meta-observables'. The non-Hermitian inner product that is used
in the dS case is based on the $\cal CPT$ theorem in the same way
that the $\cal CPT$ inner product is used in $\cal PT$-symmetric
quantum theory \cite{BBJ}.

\vspace{0.5cm}

\begin{footnotesize}
\noindent We thank Professor M Znojil for hospitality of the Czech
Academy of Sciences, where this work originated. CMB is grateful
to the US Department of Energy for financial support. MVB's
research is supported by the Royal Society of London.
\end{footnotesize}

\end{document}